\documentclass[twocolumn]{aastex631}
\usepackage{amsmath,amssymb,amstext}
\usepackage{gensymb}
\usepackage{multirow}
\usepackage{rotating}
\usepackage{tabularx}
\usepackage{comment}
\usepackage{multirow}
\usepackage{natbib}

\usepackage{xspace}

\newcommand{\jwst}{\texttt{jwst}\xspace}
\newcommand{\eureka}{\texttt{Eureka!}\xspace}

\newcommand{\jedi}{\texttt{ExoTiC-JEDI}\xspace}
\newcommand{\picaso}{\texttt{PICASO}\xspace}

\defcitealias{Lustig-Yaeger2023}{Lustig-Yaeger \& Fu et al. 2023}
\defcitealias{Moran2023}{Moran \& Stevenson et al. 2023}
\defcitealias{May2023}{May \& MacDonald et al. 2023}

\begin{document}

\title{JWST COMPASS: NIRSpec/G395H Transmission Observations of the Super-Earth TOI-776b}


\author[0000-0001-8703-7751]{Lili Alderson} 
\affiliation{Department of Astronomy, Cornell University, 122 Sciences Drive, Ithaca, NY 14853, USA}
\affiliation{School of Physics, University of Bristol, HH Wills Physics Laboratory, Tyndall Avenue, Bristol BS8 1TL, UK}

\author[0000-0002-6721-3284]{Sarah E. Moran} 
\altaffiliation{NHFP Sagan Fellow}
\affiliation{NASA Goddard Space Flight Center, 8800 Greenbelt Road, Greenbelt, MD 20771, USA}


\author[0000-0003-0354-0187]{Nicole L. Wallack}
\affiliation{Earth and Planets Laboratory, Carnegie Institution for Science, 5241 Broad Branch Road, NW, Washington, DC 20015, USA}


\author[0000-0003-1240-6844]{Natasha E. Batalha}
\affiliation{NASA Ames Research Center, Moffett Field, CA 94035, USA}

\author[0000-0002-0413-3308]{Nicholas F. Wogan}
\affiliation{NASA Ames Research Center, Moffett Field, CA 94035, USA}

\author[0000-0002-1092-2995]{Anne Dattilo}
\affiliation{Department of Astronomy and Astrophysics, University of California, Santa Cruz, CA 95064, USA}


\author[0000-0003-4328-3867]{Hannah R. Wakeford} 
\affiliation{School of Physics, University of Bristol, HH Wills Physics Laboratory, Tyndall Avenue, Bristol BS8 1TL, UK}


\author[0000-0002-4489-3168]{Jea Adams Redai} 
\affiliation{Center for Astrophysics ${\rm \mid}$ Harvard {\rm \&} Smithsonian, 60 Garden St, Cambridge, MA 02138, USA}

\author[0000-0003-4157-832X]{Munazza K. Alam}
\affiliation{Space Telescope Science Institute, 3700 San Martin Drive, Baltimore, MD 21218, USA}

\author[0000-0002-8949-5956]{Artyom Aguichine}
\affiliation{Department of Astronomy and Astrophysics, University of California, Santa Cruz, CA 95064, USA}

\author[0000-0002-7030-9519]{Natalie M. Batalha}
\affiliation{Department of Astronomy and Astrophysics, University of California, Santa Cruz, CA 95064, USA}

\author[0009-0003-2576-9422]{Anna Gagnebin} 
\affiliation{Department of Astronomy and Astrophysics, University of California, Santa Cruz, CA 95064, USA}

\author[0000-0002-8518-9601]{Peter Gao} 
\affiliation{Earth and Planets Laboratory, Carnegie Institution for Science, 5241 Broad Branch Road, NW, Washington, DC 20015, USA}


\author[0000-0002-4207-6615]{James Kirk} 
\affiliation{Department of Physics, Imperial College London, London, UK}

\author[0000-0003-3204-8183]{Mercedes L\'opez-Morales} 
\affiliation{Space Telescope Science Institute, 3700 San Martin Drive, Baltimore, MD 21218, USA}

\author[0000-0002-7500-7173]{Annabella Meech}
\affiliation{Center for Astrophysics ${\rm \mid}$ Harvard {\rm \&} Smithsonian, 60 Garden St, Cambridge, MA 02138, USA}


\author[0009-0008-2801-5040]{Johanna Teske} 
\affiliation{Earth and Planets Laboratory, Carnegie Institution for Science, 5241 Broad Branch Road, NW, Washington, DC 20015, USA}
\affiliation{The Observatories of the Carnegie Institution for Science, 813 Santa Barbara St., Pasadena, CA 91101, USA}

\author[0000-0003-2862-6278]{Angie Wolfgang}
\affiliation{Eureka Scientific Inc., 2452 Delmer Street Suite 100, Oakland, CA 94602-3017}

\begin{abstract}

We present two transit observations of the $\sim$520\,K, 1.85\,R$_\oplus$, 4.0\,M$_\oplus$ super-Earth TOI-776b with JWST NIRSpec/G395H, resulting in a 2.8--5.2\,$\mu$m transmission spectrum. 
Producing reductions using the \jedi and \eureka pipelines, we obtain a median transit depth precision of 34\,ppm for both visits and both reductions in spectroscopic channels 30 pixels wide ($\sim0.02\,\mu$m). 
We find that our independent reductions produce consistent transmission spectra, however, each visit shows differing overall structure. For both reductions, a flat line is preferred for Visit 1 while a flat line with an offset between the NRS1 and NRS2 detectors is preferred for Visit 2; however, we are able to correct for this offset during our modeling analysis following methods outlined in previous literature. Using \picaso forward models, we can rule out metallicities up to at least 100$\times$ solar with an opaque pressure of 10$^{-3}$ bar to $\geq$3$\sigma$ in all cases, however, the exact lower limit varies between the visits, with Visit 1 ruling out $\lesssim$100$\times$ solar while the lower limits for Visit 2 extend beyond $\sim$350$\times$ solar. Our results add to the growing list of super-Earth atmospheric constraints by JWST, which provide critical insight into the diversity and challenges of characterizing terrestrial planets. 

\end{abstract}


\section{Introduction} 
\label{section:intro}

Super-Earth and sub-Neptune exoplanets are numerous in our Galaxy \citep[e.g.,][]{Dressing2015}. However, their existence -- along with those that exist in the radius gap in between -- presents challenges to our current understanding of planet formation and evolution \citep[e.g.,][]{Ginzburg2018, RogersOwen2021, Rogers2021b, Owen2023}, and leaves many open questions as to how diverse their interiors and atmospheres may be \citep[e.g.,][]{Heng2012,Dorn2015,Zeng2016,Brugger2017,Zahnle2017,Lichtenberg2025}. As such, detailed study of both individual super-Earths and of the population as a whole is currently seen as a major strategic goal in the field of exoplanets (e.g., Astronomy \& Astrophysics Decadal Survey\footnote{https://www.nationalacademies.org/our-work/decadal-survey-on-astronomy-and-astrophysics-2020-astro2020}; NASA Exoplanet Exploration Program Science Gap List\footnote{https://exoplanets.nasa.gov/exep/science-overview/}), and is a focus of efforts with both current instruments \citep[e.g.,][]{Seifahrt2020,Teske2021,Redfield2024} and upcoming observatories (e.g., \citealt{Kammerer2022,Currie2023,Zhang2024ELT}).  

Transmission spectroscopy presents one method of characterizing super-Earths in greater detail; however, their often sub-percent transit depths and small scale height atmospheres require precisions of order 10\,ppm to be sensitive to their atmospheres. With the era of JWST well underway, its low instrumental noise floors ($\lesssim$10\,ppm, \citealt{Schlawin2021}; \citetalias{Lustig-Yaeger2023}) have enabled the atmospheric study of super-Earths to become increasingly commonplace (e.g., \citetalias{May2023,Moran2023}; \citealt{Alam2024, Alderson2024, Cadieux2024, Damiano2024, Gressier2024, Scarsdale2024}). Thus far, JWST has typically found super-Earths to have high mean molecular weight atmospheres, with constraints such as lower metallicity limits at hundreds of times the solar value \citep[e.g.,][]{Alderson2024,Scarsdale2024}, or the data preferring an atmosphere rich in heavy elements such as sulfur \citep[e.g.,][]{Banerjee2024,Gressier2024} or nitrogen \citep[e.g.,][]{Cadieux2024, Damiano2024}. However, the effort to obtain robust inferences from these observations has not been without challenges. Detector offsets (e.g., \citetalias{Moran2023}; \citealt{Gressier2024}), discrepancies between multiple visits \citepalias[e.g.,][]{May2023}, unocculted faculae (e.g., \citetalias{May2023,Moran2023}; \citealt{Banerjee2024,Cadieux2024}) and varying detection significances between reductions \citep[e.g.,][]{Gressier2024,Kirk2024} all contribute to an uncertain picture, further muddied by transmission spectra that can be explained by a variety of atmospheric compositions \citep[e.g.,][]{Damiano2024,Scarsdale2024}.

Aimed at obtaining atmospheric constraints for a statistically motivated sample of 1--3\,R$_{\oplus}$ planets, the JWST COMPASS (Compositions of Mini-Planet Atmospheres for Statistical Study) Program (GO-2512, PIs N. E. Batalha \& J. Teske) is using NIRSpec/G395H to explore the broader small exoplanet population. In total, the program will assess 12 planets, with four pairs in the same systems (for full details regarding the COMPASS program and its sampling method, see \citealt{Batalha2023, Alderson2024, Wallack2024})\footnote{Our full sample includes L\,98-59\,c and L\,98-59\,d, the latter of which is being observed by GTO-1224, PI S. Birkmann.}. Previous results from COMPASS include: ruling out an H$_2$-dominated atmosphere with a lower limit on its atmospheric metallicity of 250$\times$ solar for the super-Earth TOI-836b \citep{Alderson2024}; finding that atmospheres both $>$175$\times$ solar and lower metallicity atmospheres with highly lofted clouds are plausible for the sub-Neptune TOI-836c \citep{Wallack2024}; ruling out $<$300$\times$ solar and pure methane atmospheres for the super-Earth L\,98-59\,c \citep{Scarsdale2024}; ruling out metallicities $<$100$\times$ solar for a cloudless atmosphere scenario for the rocky L\,168-9\,b \citep{Alam2024}; and ruling out metallicities $<$180-240$\times$ solar for the water world candidate TOI-776c (Teske \& Batalha et al., submitted).

Continuing in COMPASS' exploration is TOI-776b (TOI-776.02), the inner planet of the multi-planet TOI-776 (LP\,961-53) system. TOI-776, a bright (J mag $\sim$ 8.48) M-dwarf \citep{Luque2021}, is also host to TOI-776c (TOI-776.01), another COMPASS target (Teske \& Batalha et al., submitted). The subject of this paper, TOI-776b, orbits on a period of 8.25\,days, with a radius of 1.85$\pm$0.13\,R$_{\earth}$, and a mass of 4.0$\pm$0.9\,M$_{\earth}$  \citep[][see Table \ref{table:system} for detailed system parameters]{Fridlund2024}. 

This paper is laid out as follows: In Section \ref{section:obs}, we detail our observations and describe our reduction methods in Section \ref{section:data_reduction}. We present the transmission spectrum of TOI-776b in Section \ref{section:results}, and interpret the transmission spectrum using both agnostic statistical methods and 1D radiative-convective atmospheric models in Section \ref{section:interpretation}. Finally, we discuss our interpretations and summarize our conclusions in Section \ref{section:discussion}. 

\section{Observations} 
\label{section:obs}

We observed two transits of TOI-776b with JWST NIRSpec using the high-resolution (R$\sim$2700) G395H mode, commencing on May 24 2023 at 02:56 UTC and June 9 2023 at 14:48 UTC. We note that a High Gain Antenna (HGA) move occurred during the first transit, approximately 5 hours into the observation, shortly before egress (see Section \ref{section:data_reduction}). The NIRSpec/G395H mode results in spectroscopy from 2.87--5.14\,$\mu$m, split across the NRS1 and NRS2 detectors (with a gap between 3.72--3.82\,$\mu$m). For both observations, we used the NIRSpec Bright Object Time Series (BOTS) mode with the SUB2048 subarray, F290LP filter, S1600A1 slit, and NRSRAPID readout pattern.  Both observations lasted 6.6 hours and consisted of 3288 integrations with 7 groups per integration, for an effective integration time of 6.3\,s. Each observation also consisted of sufficient time to cover the 1.8-hour transit as well as pre- and post-transit baseline.

\begin{deluxetable}{lll}

\tablewidth{0pt}
\tablehead{\colhead{Property}&\colhead{\citet{Luque2021}}&\colhead{\citet{Fridlund2024}}}
\startdata
K (mag)\tablenotemark{a}  & 7.615 $\pm$ 0.02& 7.615 $\pm$ 0.02\\
$R_*$ (R$_\odot$)& 0.538 $\pm$ 0.024 &  0.547 $\pm$ 0.017\\
T$_*$ (K)& 3709 $\pm$ 70 & 3725 $\pm$ 60\\
log(g) & 4.727 $\pm$ 0.025 &  4.8 $\pm$ 0.1 \\
$[$Fe/H$]_{*}$& -0.20 $\pm$ -0.12 & -0.21 $\pm$ 0.08 \\
\hline
Period (days)& 8.24661$^{+0.00005}_{-0.00004}$ & 8.246620$^{+0.000024}_{-0.000031}$\\
$M_\mathrm{P}$ (M$_{\earth}$)&  4.0 $\pm$ 0.9 & 5.0 $\pm$ 1.6\\ 
$R_\mathrm{P}$ (R$_{\earth}$)& 1.85 $\pm$ 0.013 & 1.798$^{+0.078}_{-0.077}$\\
T$_\mathrm{eq}$ (K)& 514 $\pm$ 17 & 520 $\pm$ 12\\
$e$& 0.06$^{+0.03}_{-0.02}$ & 0.052$^{+0.037}_{-0.035}$\\
$\omega$ (\textdegree)& -67$^{+117}_{-73}$ & 45$^{+94}_{-110}$
\tablecaption{System Properties for TOI-776b. Values used in the light curve fitting are shown in Table \ref{table:fit_table}.}
\enddata 
\tablenotetext{a}{Both values from 2MASS} 
\label{table:system}
\end{deluxetable}

\section{Data Reduction} 
\label{section:data_reduction}

Following other NIRSpec/G395H transmission spectra analyses, we reduced the data using two independent pipelines to assess the robustness of our resulting conclusions. The details of our \texttt{ExoTiC-JEDI} \citep{Alderson2022, Alderson2023} and \texttt{Eureka!} \citep{Bell2022} reductions are described below (see Sections \ref{section:jedi} and \ref{sec:eureka} respectively).  

\subsection{ExoTiC-JEDI} \label{section:jedi}

The Exoplanet Timeseries Characterisation - JWST Extraction and Diagnostic Investigator (\jedi) package \citep{Alderson2022JEDI}\footnote{https://github.com/Exo-TiC/ExoTiC-JEDI} is designed to perform an end-to-end extraction, reduction, and analysis of JWST time-series data from raw \texttt{uncal} files through to light curve fitting to produce planetary spectra. Throughout the analysis, each visit is treated separately with data from NRS1 and NRS2 reduced independently, testing a variety of values for each reduction parameter. The resulting reduction setup as described below uses the values that provided the smallest out-of-transit scatter in the resulting white light curve.

We performed our \jedi reduction with the standard set-up as described in \citet{Alderson2024}, beginning with Stages 1 and 2, which operate as a modified version of Stage 1 of the \jwst pipeline \citep[v.1.14.0, context map 1225;][]{Bushouse2022}. In Stage 1 we carried out custom bias subtraction; group level 1/$f$ noise destriping, masking the spectral trace 15$\sigma$ from the dispersion axis for each integration; linearity, dark current, and saturation corrections; jump detection, using a threshold of 15$\sigma$; and ramp fitting before producing 2D wavelength maps in Stage 2 to obtain the wavelength solutions. In Stage 3, the reduction corrected for the `do not use', `saturated', `dead', `hot', `low quantum efficiency', and `no gain value' data quality flags, and replaced any other pixels that were greater than 6$\sigma$ outliers spatially with the median of the neighboring 4 pixels, and 20$\sigma$ outliers temporally with the median of the nearest 10 pixels. Column-by-column destriping was repeated to remove any remaining 1/$f$ and background before the location of the spectral trace was obtained. For both NRS1 and NRS2, the data reduction process favored an aperture five times the FWHM of the spectral trace, approximately 8 pixels wide from edge to edge. As with previous \jedi data reductions, an intrapixel extraction was used \citep[e.g.,][]{Alderson2024, Wallack2024}. Alongside extracting the 1D stellar spectra, $x$- and $y$-positional shifts were measured using the 1D stellar spectra, for use in systematic light curve detrending.

\begin{table*}[ht!]
\centering
\caption{Best fit values for the four individual white light curve fits for \jedi and \eureka as shown in Figure \ref{figure:wlc}.}
\label{table:fit_table}
\begin{tabular}{ccc|c|c|c|c}
\multicolumn{3}{c|}{}                     & T$_{0}$ (MJD) & $a/R_*$ & $i$ ($\degree$) & R$_p$/R$_*$ \\ \hline
\multicolumn{3}{c|}{\citet{Luque2021}}           & 58571.4167$\pm$1e-3  &  27.87$^{+0.97}_{-1.02}$  &  89.65$^{+0.22}_{-0.37}$ & 0.0316$^{+0.0008}_{-0.0011}$  \\ \hline
\multicolumn{3}{c|}{\citet{Fridlund2024}}           & 59288.8713$^{+1.0e-3}_{-1.1e-3}$  &  --   &  89.41$^{+0.39}_{-0.36}$ & --   \\ \hline
\multicolumn{1}{c|}{\multirow{4}{*}{\jedi}}   & \multicolumn{1}{c|}{\multirow{2}{*}{Visit 1}} & NRS1 & 60088.29193$\pm$5e${-5}$  & 28.64$\pm$1.25   & 89.72$\pm$0.55 & 0.03075$\pm$0.00012  \\
\multicolumn{1}{c|}{}                                        & \multicolumn{1}{c|}{}                         & NRS2 & 60088.29193 $\pm$7e${-5}$   & 26.30$\pm$1.22   & 89.13$\pm$0.23 & 0.03068$\pm$0.00013  \\
\multicolumn{1}{c|}{}                                        & \multicolumn{1}{c|}{\multirow{2}{*}{Visit 2}} & NRS1 & 60104.78585$\pm$6e${-5}$ & 28.69$\pm$1.36   & 89.65$\pm$0.49 & 0.02980$\pm$0.00013  \\
\multicolumn{1}{c|}{}                                        & \multicolumn{1}{c|}{}                         & NRS2 & 60104.78573$\pm$7e${-5}$   & 26.36$\pm$1.16   & 89.14$\pm$0.22 & 0.03096$\pm$0.00012  \\ \hline
\multicolumn{1}{c|}{\multirow{4}{*}{\eureka}} & \multicolumn{1}{c|}{\multirow{2}{*}{Visit 1}} & NRS1 & 60088.29198 $\pm 5$e${-5}$         & 28.66$\pm$0.50 & 89.75$\pm$0.19         & 0.03088$\pm$0.00009 \\
\multicolumn{1}{c|}{}                                        & \multicolumn{1}{c|}{}                         & NRS2 & 60088.29197 $\pm 7$e${-5}$           & 28.17$\pm$0.96   & 89.55$\pm$0.27 & 0.03057$\pm$0.00011 \\
\multicolumn{1}{c|}{}                                        & \multicolumn{1}{c|}{\multirow{2}{*}{Visit 2}} & NRS1 & 60104.78592$\pm$5e${-5}$           & 28.63$\pm$0.44 & 89.77$\pm$0.18 & 0.02995$\pm$0.00009 \\
\multicolumn{1}{c|}{}                                        & \multicolumn{1}{c|}{}                         & NRS2 & 60104.78573$\pm$7e${-5}$           & 27.23$\pm$1.07   & 89.32$\pm$0.27 & 0.03102$\pm$0.00012 \\
\end{tabular}
\end{table*}

\begin{figure*}
\begin{centering}
\includegraphics[width=0.9899\textwidth]{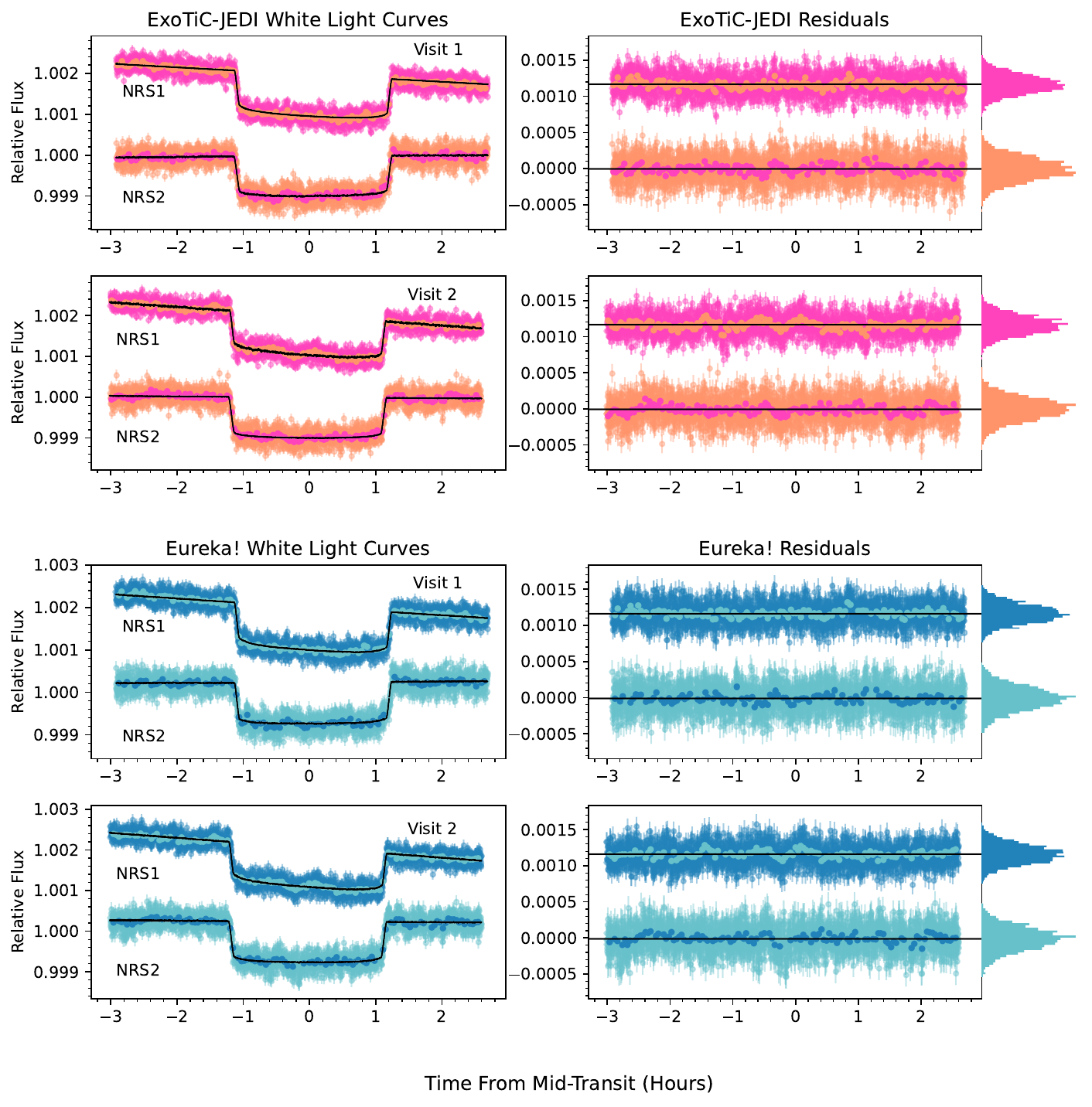}
\caption{The \jedi and \eureka white light curves for each detector and visit, with best-fit models and the associated residuals. The alternate colors demonstrate binned light curves and residuals, which are plotted for reference. Histograms of the residuals are shown in the rightmost column.} 
\label{figure:wlc}   
\end{centering}
\end{figure*} 

We fit white light curves for both NRS1 and NRS2, as well as spectroscopic light curves across the full NIRSpec/G395H wavelength range. White light curves span from 2.814--3.717\,$\mu$m for NRS1 and 3.824--5.111\,$\mu$m for NRS2, while spectroscopic light curves were each 30-pixels wide ($\sim0.02$\,\micron, $R\sim200$). For the white light curves, we fit for the system inclination, $i$, the ratio of the semi-major axis to the stellar radius, $a/R_*$, the center of transit time, $T_0$, and the ratio of the planet to stellar radii, $R_p/R_*$, holding the period, $P$, eccentricity, $e$, and argument of periastron, $\omega$, fixed to the values presented in \citet{Luque2021}\footnote{We elected to use \citet{Luque2021} for our starting values as \citet{Fridlund2024} do not report values for $a/R_*$ or $R_p/R_*$, and $P$, $e$ and $\omega$ are consistent between the two studies within the reported uncertainties.}. The stellar limb darkening coefficients are held fixed to values calculated using the \texttt{ExoTiC-LD} package \citep{Grant2024} based on the stellar T$_{*}$, log(g), and [Fe/H]$_{*}$ presented in \citet{Luque2021} (see Table \ref{table:system}), with Phoenix stellar models \citep{Husser2013} and the non-linear limb darkening law \citep{Claret2000}.

We used a least-squares optimizer to fit for a \texttt{batman} \citep{Kreidberg2015} transit model simultaneously with our systematic model $S(\lambda)$, which took the form
$$     S(\lambda) = s_0 + (s_1 \times x_s) + (s_2 \times y_s) + (s_3 \times t) \mathrm{,}$$
where $x_s$ is the $x$-positional shift of the spectral trace, $y_s$ is the $y$-positional shift of the spectral trace, $t$ is the time and $s_0, s_1, s_2$ and $s_3$ are coefficient terms. For the spectroscopic light curves, we fit for $R_p/R_*$, holding $T_0$, $i$, and $a/R_*$ fixed to the respective white light curve fit value, as shown in Table \ref{table:fit_table}. For both the white and spectroscopic light curves, we removed any data points that were greater than 4$\sigma$ outliers in the residuals and refit the light curves until no such points remained. We also trimmed the first 500 integrations ($\sim$60 minutes) to remove any initial persistence or settling ramp from the detector. Due to the presence of the aforementioned HGA move in Visit 1, we additionally removed a total of 15 integrations surrounding the timing of the move that appeared to have large $y$-positional shifts. We also rescaled the flux time series errors using the beta value \citep{Pont2006} as measured from the white and red noise values calculated using the \texttt{noise\_calculator()} in \jedi to account for any remaining red noise in the data. The fitted white light curves and residuals for \jedi are shown in Figure \ref{figure:wlc}, while the best-fit parameters are listed in Table \ref{table:fit_table}. Spectroscopic light curves are shown in the Appendix (Figure \ref{figure:2d_plot}).

\subsection{Eureka!}\label{sec:eureka}

We produce a second independent reduction using the open-source \texttt{Eureka!} package \citep{Bell2022}\footnote{https://eurekadocs.readthedocs.io/en/latest/index.html}, an end-to-end pipeline for the reduction of JWST and HST data, in the same way as described in \cite{Alderson2024}. For the reduction of JWST data, \eureka acts as a wrapper around the first two stages of the \texttt{jwst} pipeline (v.1.11.4, context map jwst\_1225.pmap). We use the default Stage 1 inputs for the \texttt{jwst} pipeline with the exception of using a jump detection threshold of 15$\sigma$. \eureka also performs custom group-level background subtraction in Stage 1 to account for the 1/$f$ noise, removing a column-by-column median from the trace-masked image. For Stage 2, we utilize the default \texttt{jwst} pipeline steps within \eureka. For Stage 3, we then optimize the choice of extraction apertures (half-widths between 4--8 pixels), background apertures (half-widths between 8--11 pixels), sigma thresholds for optimal extraction outlier rejection, and polynomial order for an additional background subtraction by selecting the combination of reduction parameters that minimizes the median absolute deviation of the white light curves. For Visit 1, we find optimal apertures of 5 pixels for both NRS1 and NRS2 and background apertures of 11 and 8 pixels for NRS1 and NRS2 respectively. For Visit 2, we find optimal extraction apertures of 4 pixels and background apertures of 8 pixels for both detectors. 

We then extract 30-pixel wide spectroscopic light curves, to be fit following the white light curves (2.863--3.714\,$\micron$ for NRS1 and 3.812--5.082\,$\micron$ for NRS2). We do not use the inbuilt \eureka light curve fitter, opting for the same fitter described in \cite{Alderson2024} for increased flexibility. For both our spectroscopic and white light curves, we first iteratively trim 3$\sigma$ outliers from a 50-point rolling median three times. We use the Markov chain Monte Carlo (MCMC) package \texttt{emcee} \citep{Foreman-Mackey2013} to simultaneously fit a transit and systematic model to the light curves. For the white light curves, we fit for $i$, $a/R_{*}$, $T_{0}$, and $R_{p}/R_{*}$ with uninformed priors using \texttt{batman} \citep{Kreidberg2015}, fixing the quadratic limb-darkening coefficients to the theoretical values computed with Set One of the MPS-ATLAS models \citep{Kostogryz2022} with \texttt{ExoTiC-LD} \citep{Grant2024} using stellar parameters from \citet{Luque2021} (see Table \ref{table:system}). We fix the period and eccentricity to the values in Table~\ref{table:system}. Our systematic model, $S(\lambda)$, took the form
\begin{equation}
S(\lambda) = s_{0} + (s_{1} \times x)+ (s_{2}\times y) + (s_{3}\times t) , 
\label{eq:1}
\end{equation}
where $s_{N}$ are parameters fitted for in our instrumental noise model, $t$ is the timing array, and $x$ and $y$ are the positions of the trace in each detector. We note that this systematic model is the same form as that used for the \jedi reduction. In addition to the systematic model and astrophysical model, we fit for an additional per-point error term that is added in quadrature to the measured errors. We trim the first 500 points ($\sim$60 minutes) of the observation for both visits to remove any initial ramp that may bias the slope of the systematic noise model. 

\begin{figure*}
\begin{centering}
\includegraphics[width=0.975\textwidth]{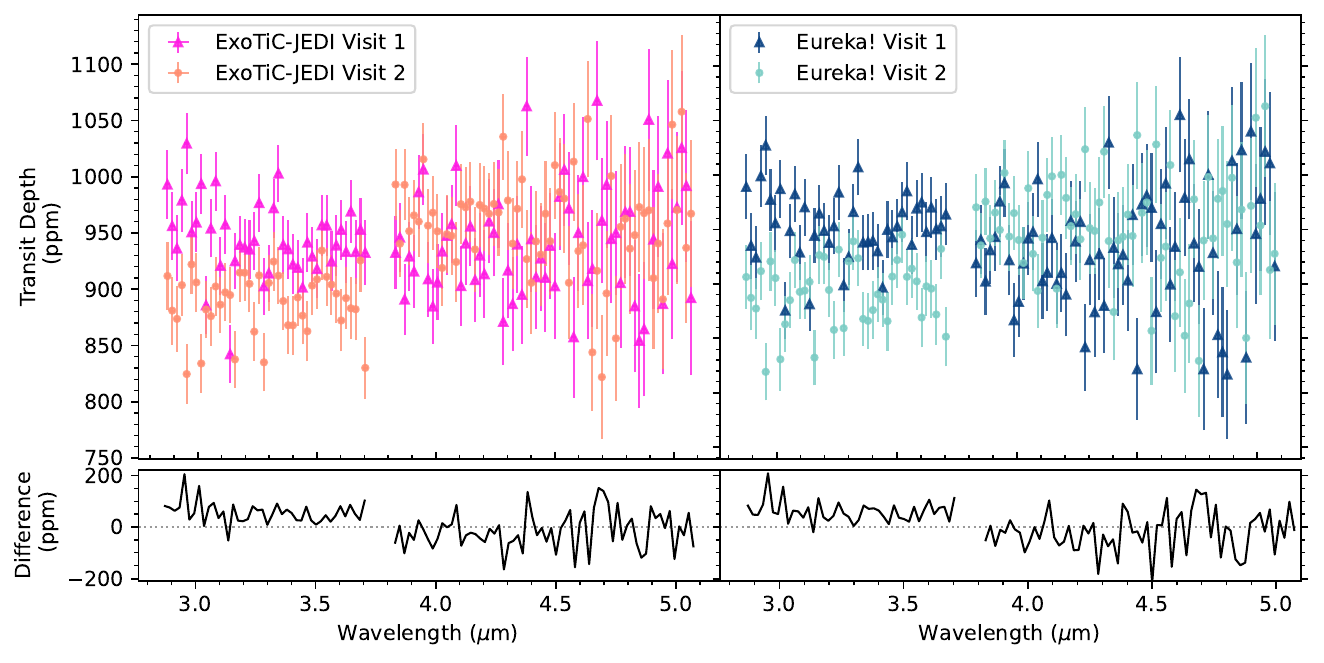}
\includegraphics[width=0.975\textwidth]{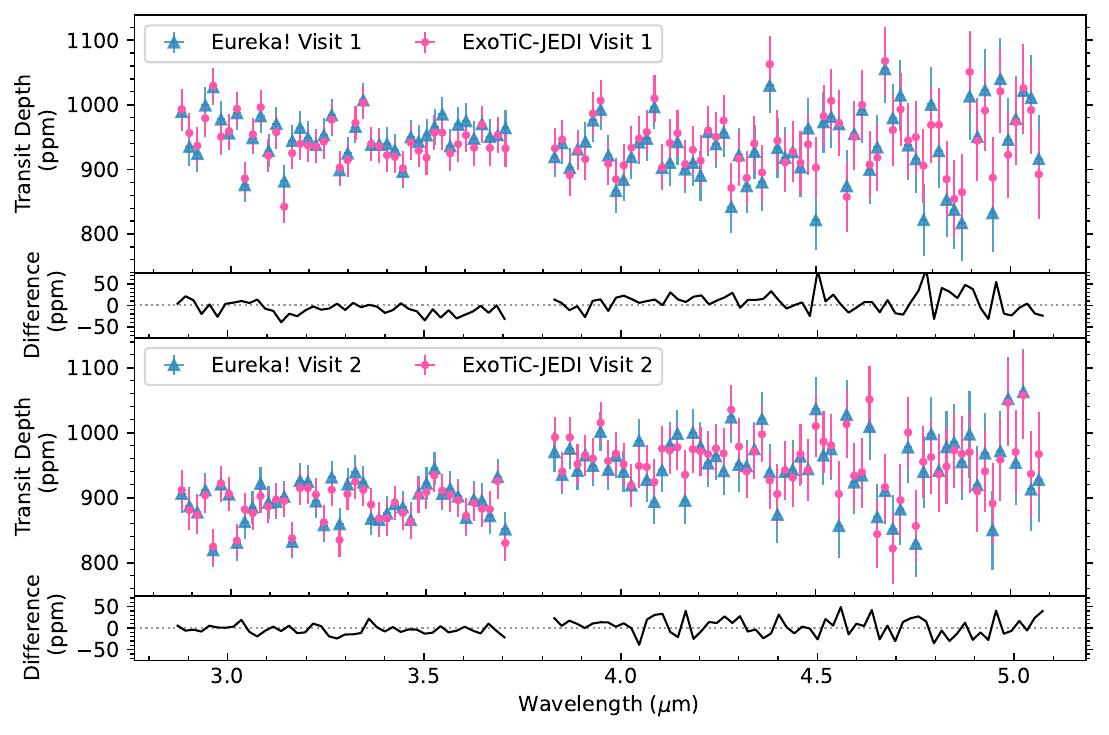}
\caption{Top: Transmission spectra produced by the \jedi (left) and \eureka (right) reductions, with the difference between the two visits shown by the solid line in the lower panel. Across both pipelines, there is a noticeable difference in the morphology between the two visits, particularly due to the transmission spectra of NRS1 in Visit 2 being offset relative to NRS2. Bottom: Transmission spectra from Visit 1 (upper panels) and Visit 2 (lower panels) comparing the two different pipelines. \jedi is shown by the pink circles and \eureka is shown by the blue triangles. The difference between the two reductions for each visit is shown by the solid line plotted beneath each transmission spectrum. The pipelines have a median difference of 13\,ppm for Visit 1 and 12\,ppm for Visit 2 compared to a median transit depth uncertainty of 34\,ppm for both Visit 1 and 2.} 
\label{figure:tspec}   
\end{centering}
\end{figure*} 

We first run a Levenberg-Marquardt minimization to get an initial guess for the free parameters. We set the number of MCMC walkers equal to  3$\times$ the number of free parameters. We fix the best fit $i$, $a/R_{*}$ and $T_{0}$ from the white light curves for each detector in the subsequent spectroscopic light curve fits, where we fit for $R_{p}/R_{*}$, utilizing the same functional form of the systematic noise model and the same fitting procedure as employed for the white light curves. The fitted white light curves and residuals for \eureka are shown in Figure \ref{figure:wlc}, while the best-fit parameters are listed in Table \ref{table:fit_table}. Spectroscopic light curves are shown in the Appendix (Figure \ref{figure:2d_plot}).


\section{Transmission Spectrum} 
\label{section:results}

\begin{figure}
\begin{centering}
\includegraphics[width=0.45\textwidth]{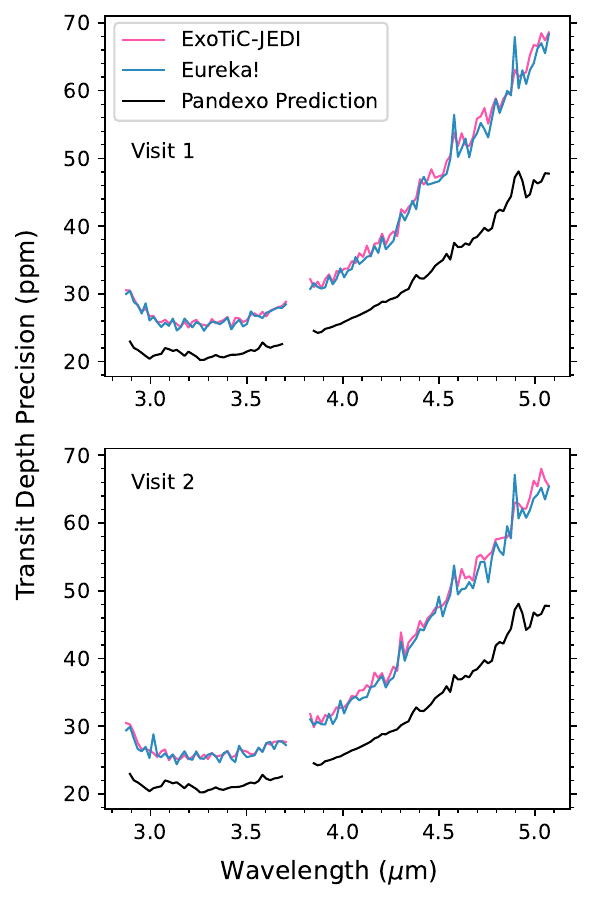}
\caption{Transit depth precisions achieved by the \jedi (pink) and \eureka (blue) reductions compared to predicted values from \texttt{PandExo} simulations for the observation set-up. As seen in other low group number studies (e.g., \citetalias{Lustig-Yaeger2023, Moran2023}; \citealt{Alderson2024, Wallack2024}) the observations do not match the precision expected by \texttt{PandExo}. They are on average 1.3$\times$ the prediction, with all values within 1.5$\times$ the \texttt{PandExo} value. We obtain a median transit depth precision in wavelength bins 30 pixels wide ($\sim0.02$\,\micron, $R\sim200$) of 34\,ppm for both reductions of both visits.}
\label{figure:pandexo}   
\end{centering}
\end{figure} 

The 3--5\,$\mu$m transmission spectra of TOI-776b using a 30-pixel binning scheme for the \jedi and \eureka reductions are shown at the top of Figure \ref{figure:tspec}, where no offsets have been applied between NRS1 and NRS2 or between the visits. For both visits and both reductions, the median transit depth uncertainty is 34\,ppm. For both reductions, there is a noticeable difference between the overall structure of the transmission spectrum in each visit, with Visit 1 appearing broadly flat across the full wavelength range, while for Visit 2, the transmission spectrum across the NRS1 detector appears to be offset relative to the NRS2 detector (for further discussion see Section \ref{section:synth_fits}). For this reason, we elected not to produce a combined visit spectrum before investigating whether our model interpretation differed per visit. As discussed in Section \ref{section:interpretation}, we saw significant differences in the limits between each visit, and therefore we do not produce a combined visit spectrum. As shown in the bottom panel of Figure \ref{figure:tspec}, both reductions are consistent between each visit - i.e., both \jedi and \eureka show very similar morphologies for Visit 1 and Visit 2 respectively, with a median difference between the transit depths of 13\,ppm for Visit 1 and 12\,ppm for Visit 2. 

In Figure \ref{figure:pandexo}, we compare the transit depth precisions from \jedi and \eureka for each visit to the precisions predicted by \texttt{PandExo} \citep{Batalha2017}. While both reductions achieve comparable precisions, neither reaches the values predicted by \texttt{PandExo} across all wavelengths, with the maximum \jedi and \eureka precisions both 1.5$\times$ the \texttt{PandExo} value. On average, both reductions are within 1.3$\times$ the \texttt{PandExo} value. This result is consistent with that seen of other NIRSpec transiting exoplanet programs around bright targets, including others within COMPASS (e.g., \citetalias{Lustig-Yaeger2023, Moran2023}; \citealt{Alderson2024,Wallack2024}).


\section{Interpretation of TOI-776\MakeLowercase{b's} Atmosphere} 
\label{section:interpretation}

\begin{table*}[ht!]
\centering
\caption{Results of non-physical fits to Visit 1 \& 2 of both \jedi and \eureka data reductions. The parameter fit column signifies: for the zero-slope model, the $(R_p/R_*)^2$ baseline intercept in ppm units, 2) for the step function model, the offset between NRS1 and NRS2 in $(R_p/R_*)^2$ ppm units, 3) for the slope case, the gradient of the slope (ppm/$\mu$m), 4) for the Gaussian in NRS1 and 5) in NRS2, the central wavelength of the ``feature'' in $\mu$m.}
\begin{tabular}{l|ccccc}
                    & \multicolumn{5}{c}{\textbf{\jedi (v1/v2)}}            \\ \hline 
\textbf{Model Type} &  $\ln$Z                & $\chi^2/N$ & Key Parameter & Parameter Fit & No. Free Parameters \\ \hline
Zero Slope & -70.0/-107.0 & 1.24/1.94 &  Baseline Intercept [ppm] & 941.6$\pm$3.2/ 914.9$\pm$3.3 & 1   \\
Offset NRS1/NRS2  &  -73.0/-54.0 & 1.23/0.88 & NRS1/NRS2 Offset [ppm] & -3.5$\pm$6.7/ 70.7$\pm$7.2 & 2  \\
Slope & -73.0/-70.0 & 1.23/1.19 & Slope Gradient [ppm/$\mu$m] & -3.54$\pm$5.87/51.89$\pm$5.6 & 2 \\
Gaussian NRS1 &  -74.0/-55.0 & 1.23/0.88 & $\lambda_0$ [$\mu$m] & 3.33$\pm$0.23/3.43$\pm$0.21 & 5 \\
Gaussian NRS2 &  -73.0/-54.0 & 1.23/0.88 & $\lambda_0$ [$\mu$m]  & 4.62$\pm$0.4/4.36$\pm$0.44 & 5\\
 \hline \hline
                    & \multicolumn{5}{c}{\textbf{\eureka (v1/v2})}           \\ \hline
\textbf{Model Type} &  $\ln$Z  & $\chi^2/N$ & Key Parameter & Parameter Fit & No. Free Parameters \\ \hline
Zero Slope &  -79.0/-105.0 & 1.41/1.9 & Baseline Intercept [ppm] & 944.0$\pm$3.4/ 915.6$\pm$3.2  & 1 \\
Offset NRS1/NRS2  & -78.0/-66.0 & 1.33/1.11   & NRS1/NRS2 Offset [ppm] & -19.9$\pm$6.6/ 60.2$\pm$6.2 & 2  \\
Slope & -80.0/-77.0 & 1.36/1.3 & Slope Gradient [ppm/$\mu$m] & -13.48$\pm$5.82/45.45$\pm$5.59 & 2 \\
Gaussian NRS1 &  -79.0/-67.0 & 1.33/1.1   & $\lambda_0$ [$\mu$m]  & 3.39$\pm$0.23/3.45$\pm$0.2 & 5\\ 
Gaussian NRS2 &  -78.0/-67.0 & 1.33/1.11 & $\lambda_0$ [$\mu$m] & 4.62$\pm$0.42/4.43$\pm$0.42 & 5 \\
\hline \hline

\end{tabular}
\label{tab:fits}
\end{table*}

Following previous COMPASS analyses \citep{Alam2024,Alderson2024,Scarsdale2024,Wallack2024}, we use two different methods to interpret TOI-776b's atmosphere. In Section  \ref{section:synth_fits}, we perform simple statistical, non-physical model fits to the data to understand: 1) how well the data is fit by a flat or sloped line, 2) the size of any offsets between NRS1 and NRS2, 3) whether the data favors the presence of any Gaussian-shaped features, and 4) whether these fits depend on the reduction method and if they are consistent between individual visits. In Section  \ref{section:model_fits} we assess more physically motivated forward models, using \texttt{PICASO} \citep{Batalha2019, Mukherjee2023} to explore the region of metallicity and cloud top or surface pressure parameter space we can effectively rule out.

\subsection{Non-Physical Fits to the Data} \label{section:synth_fits}

\begin{figure*}
\begin{centering}
\includegraphics[width=\textwidth]{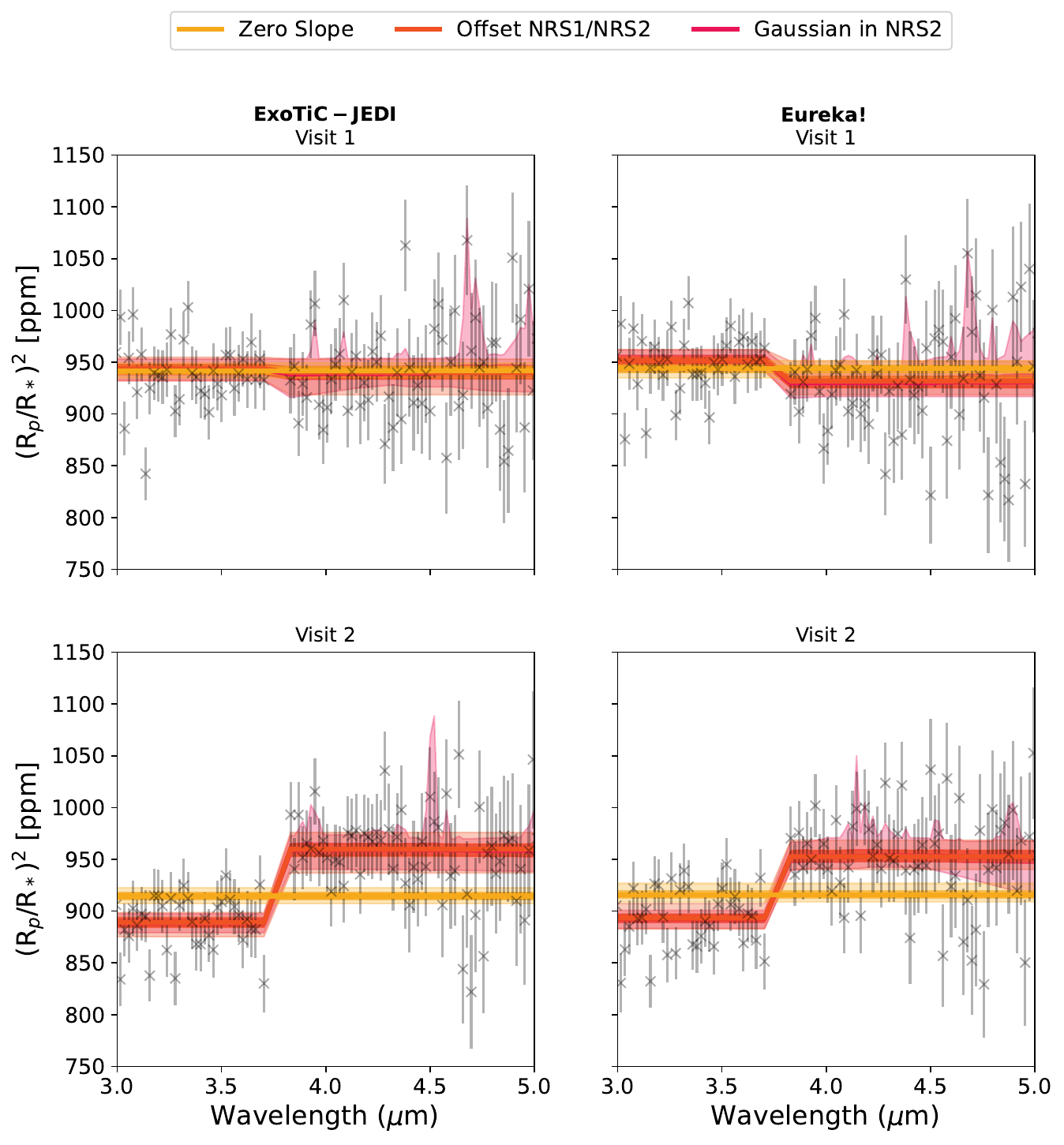}
\caption{Statistical, non-physical model fits to the data for Visit 1 (top row) and Visit 2 (bottom) for the \jedi (left column) and \eureka (right column) data reductions. Here we show the three highest-likelihood models between the two reductions. Data is shown as grey crosses; each colored line shows a different non-physical model with the shading indicating the 1$\sigma$ (dark) and 3$\sigma$ (light) regions. For Visit 1, a zero-sloped (i.e., flat) line is strongly preferred by both reductions, while an offset between NRS1 and NRS2 is preferred for Visit 2. Neither Gaussian model is preferred in each Visit.} 
\label{figure:linefits}   
\end{centering}
\end{figure*}

For the non-physical model fits, we use the \texttt{MLFriends} statistic sampler \citep{MLFriends2016, MLFriends2019} implemented in the open source code \texttt{UltraNest}  \citep{Ultranest}. For each visit and each data reduction, we fit: 1) a one-parameter, zero-slope line, 2) a two-parameter step function composed of two zero-sloped lines, one each for NRS1 and NRS2, 3) a two-parameter sloped line, 4) a five-parameter free Gaussian in NRS1 and 5) the same in NRS2. The best-fit results are shown in Table \ref{tab:fits} and Figure \ref{figure:linefits}. Across all visits and models, almost every model obtains $\chi^2/N$ values of 0.88 -- 1.4. Only Visit 2 has a $\chi^2/N$ $>$1.5 (for the flat line model), which we discuss in detail below.

We find that Visit 1, considering both reductions, is best fit by a zero-sloped -- i.e., flat -- line, devoid of spectral features. The \jedi reduction prefers the flat line among all models (with $\Delta$lnZ $\geq$3). While this flat-line non-physical model does not have the lowest log-likelihood for the \eureka reduction, ``better'' fitting models (an offset between NRS1 and NRS2 or a Gaussian feature in NRS2) have only $\Delta$lnZ $\leq$ 1.0, which does not indicate even a weak preference for these models given the corresponding increase in model complexity (2 or 5 parameters, respectively) compared to a flat line (1 parameter).

For Visit 2, we find that again both reductions agree. However, in contrast to the flat line found for Visit 1, the best-fit model for Visit 2 is instead that of two flat lines with an offset between the NRS1 and NRS2 detectors. This is strongly preferred over both single flat and sloped lines with no offset by both the \jedi and \eureka reductions, with $\Delta$lnZ $>$10. The offset model only has  $\Delta$lnZ $\leq$1 compared to the other two non-physical models, which does not rise to even a weak preference. This includes a model with a Gaussian feature in NRS2. However, as these other models have additional parameters, they are never favored over the offset model even with an equal lnZ, similar to what we find for the Gaussian feature in NRS2 for the \eureka reduction in Visit 1. Moreover, this potential Gaussian feature in NRS2 is not consistent between visits or reductions and has considerably large error bars on the central wavelength ($\sim \pm$0.4\,$\micron$). Therefore, we do not claim to detect a consistent and statistically significant feature in the NRS2 spectrum, and find that the presence of a Gaussian is driven by the noise in the data.  

Across both visits, the transit baselines corresponding to the best fitting models agree within 1$\sigma$ between the \jedi and \eureka reductions, whether for the single flat line case (Visit 1, with a baseline of $\sim$942 ppm) or the baseline plus offset scenario (Visit 2, with a baseline of $\sim$915 ppm and an offset between detectors of $\sim$65ppm). 
This $\sim$65 ppm offset observed in Visit 2 could be because of the superbias, as differing superbias treatments have been seen to produce more consistent transit depths between the detectors (see \citetalias{Moran2023}). However, the \jedi and \eureka reductions treat the superbias differently, with \jedi using a custom pseudo-bias image calculated from the median of each detector pixel in the first group across every integration in the time series \citep[see][for details]{Alderson2023}, while \eureka uses the default superbias subtraction supplied by the \jwst pipeline, and still obtain a consistent offset value. Previous work has shown, however, that one can simply fit for this offset and obtain atmospheric constraints on the scaled transit spectrum \citepalias{Moran2023}.

\subsection{Ruling out Physical Parameter Space}
\label{section:model_fits}
Accounting for the best-fit offset in Visit 2 with the values obtained in Section \ref{section:synth_fits}, we proceed to use the \picaso \citep{Batalha2019} radiative transfer code to produce atmospheric forward models to compare to our transit spectra for both visits and both reductions, independently. Our \picaso forward models use the opacities of CH$_4$, CO, CO$_2$, H$_2$O, K, NH$_3$, Na, TiO, VO, SO$_2$, OCS, N$_2$, H$_2$, and 24 other minor species, using the Zenodo v2 database of Resampled Opacities \citep{Batalha2022}. 

Our temperature-pressure profiles are computed using the parameterization of \citet{Guillot2010} with TOI-776b's zero-albedo equilibrium temperature of 514\,K. We use this temperature-pressure profile parameterization to calculate atmospheric abundances from 1$\times$ to 1000$\times$ solar metallicity in 20 logarithmic steps with a 1$\times$ solar C/O ratio of 0.55. We use chemical equilibrium calculations and thermodynamics derived from \texttt{photochem} \citep{Wogan2023}, which uses \citet{Asplund2009} for elemental (and thus solar) abundances.

After generating our chemical equilibrium atmospheres, we additionally compute a series of opaque pressure level models ranging from 1 to 10$^{-4}$ bars. These pressure levels represent either the surface of the planet or an opaque cloud deck at this pressure. Since any such cloud is fully opaque (with an optical depth $\tau$ set to 10), this cloud is fully wavelength-independent and not tied to a specific composition, acting as an agnostic cloud layer post-processed to our chemical equilibrium atmospheres.

With all chemically consistent atmospheres computed, we calculate a reduced-$\chi^2$ of the model rebinned to the resolution of the data, compared to each transmission spectrum per reduction and per visit. We treat our reduced-$\chi^2$ as a reduced-$\chi^2/N$, where $N$ is the number of data points in each spectrum. As discussed above, we account for a step function offset in the data between detectors before calculating our fits. With $\chi^2/N$ in hand, we can compute confidence intervals on our fits using \texttt{scipy.stats.distributions.chi2} and \texttt{scipy.stats.norm.ppf}, as with previous COMPASS studies (\citealt{Alderson2024,Wallack2024}, Teske \& Batalha et al., submitted) and other JWST programs \citepalias[e.g.,][]{Moran2023,May2023}.

\begin{figure*}
\begin{centering}
\includegraphics[width=0.99\textwidth]{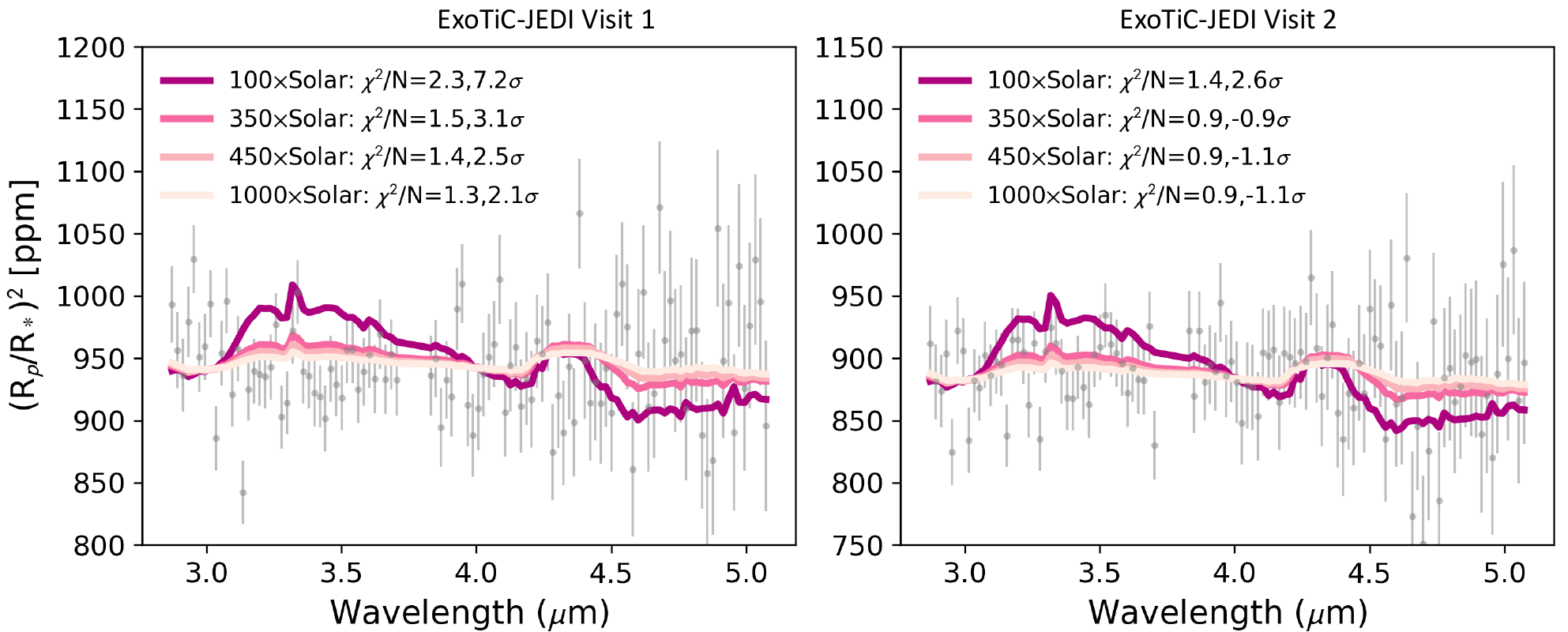}
\includegraphics[width=0.99\textwidth]{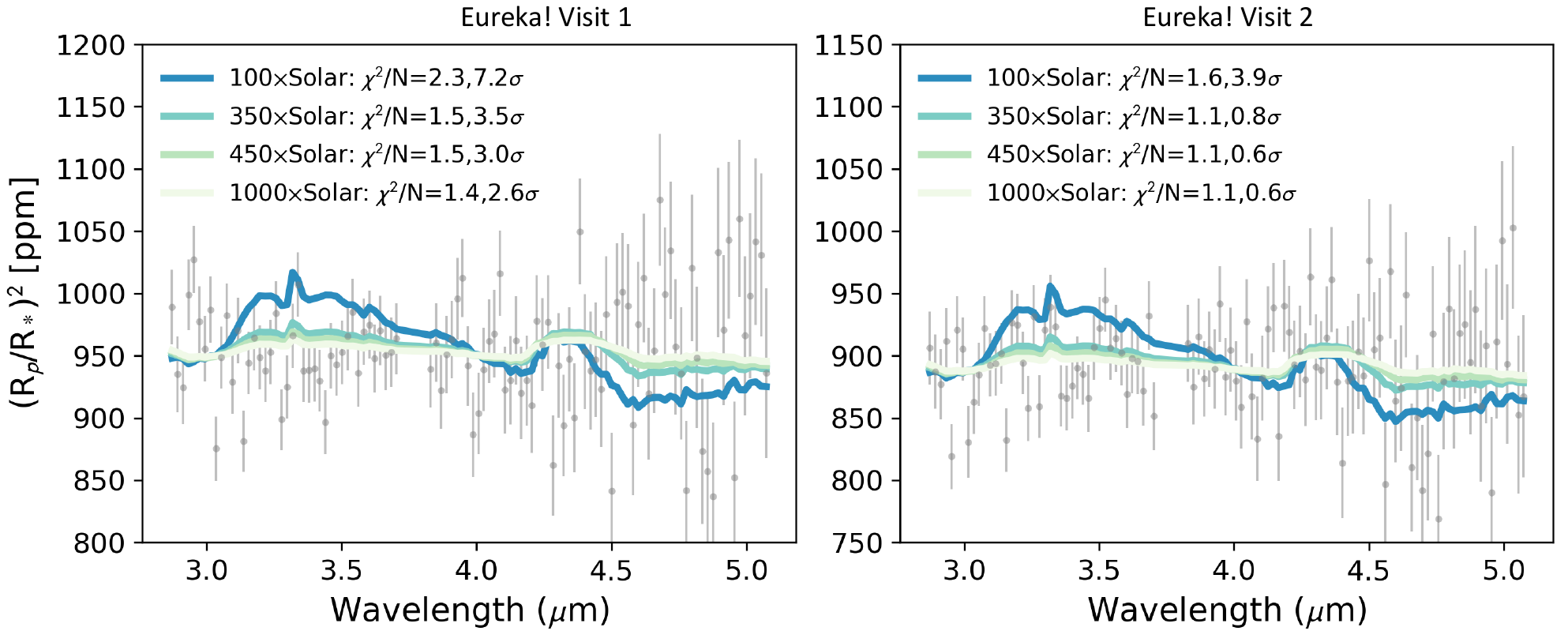}
\caption{Forward model fits to the \jedi (top, pinks) and \eureka (bottom, blues) transmission spectra for Visits 1 (left) and 2 (right), accounting for the offset in Visit 2 (see Section \ref{section:synth_fits}). Note the differing y-axis limits between visits due to differing baselines (see Table \ref{tab:fits}). We show models from 100--1000$\times$ solar metallicity for each visit with a putative 1 bar opaque pressure. The $\chi^2/N$ and confidence (in $\sigma$) by which we can rule out each model is listed next to each model. 
There are slight differences in the model limits achieved between the \jedi and \eureka reductions, with the \eureka reduction able to rule out higher metallicities in both visits.} 
\label{figure:forwardmodels}   
\end{centering}
\end{figure*}

We show a sampling of the individual results of our fits in Figure \ref{figure:forwardmodels} and Table \ref{tab:fit_table_physical}, showing 100$\times$, 350$\times$, 450$\times$, and 1000$\times$ solar metallicity atmospheres with a 1 bar opaque pressure level compared to each reduction and each visit. We strongly (by over 8$\sigma$) rule out atmospheres of 10$\times$ solar and below for atmospheres of 1 bar opaque pressure, as expected for such flat transit spectra. 
The higher metallicity atmospheres (100$\times$ to 1000$\times$ solar in Figure \ref{figure:forwardmodels}) present a more complex picture, highlighting a discrepancy in interpretation between the reductions despite the less than 1$\sigma$ differences seen per point discussed above in Section \ref{section:results}. 



\begin{table*}[]
\centering
\caption{Results of self-consistent metallicity model fits at 1 bar opaque pressure levels for a variety of metallicity values.}
\label{tab:fit_table_physical}
\begin{tabular}{ll|cc|cc|cc|cc|cc|cc|cc}
&        & \multicolumn{2}{c|}{1$\times$ solar} & \multicolumn{2}{c|}{10$\times$ solar} & \multicolumn{2}{c|}{100$\times$ solar} & \multicolumn{2}{c|}{250$\times$ solar} & \multicolumn{2}{c|}{350$\times$ solar} & \multicolumn{2}{c|}{450$\times$ solar} & \multicolumn{2}{c}{1000$\times$ solar} \\ 
    &        & $\chi^2/N$ & $\sigma$ & $\chi^2/N$ & $\sigma$ & $\chi^2/N$ & $\sigma$ & $\chi^2/N$ & $\sigma$ & $\chi^2/N$ & $\sigma$ & $\chi^2/N$ & $\sigma$ & $\chi^2/N$ & $\sigma$ \\ \hline
\hline
\multirow{2}{*}{Visit 1} & \jedi & 5.4 & 8.2 & 5.8 & 8.2 & 2.3 & 7.2 & 1.7 & 4.0 & 1.5 & 3.1 & 1.4 & 2.1 & 1.3 & 2.1 \\
    & \eureka & 5.4 & 8.2 & 5.8 & 8.2 & 2.3 & 7.2 & 1.7 & 4.3 & 1.5 & 3.5 & 1.5 & 3.0 & 1.4 & 2.6 \\
\hline
\multirow{2}{*}{Visit 2} & \jedi & 4.2 & 8.2 & 4.5 & 8.2 & 1.4 & 2.6 & 1.0 & -0.3 & 0.9 & -0.9 & 0.9 & -1.1 & 0.9 & -1.1 \\
     & \eureka & 4.6 & 8.2 & 4.8 & 8.2 & 1.6 & 3.9 & 1.2 & 1.3 & 1.1 & 0.8 & 1.1 & 0.6 & 1.1 & 0.6 \\
\hline
\hline
\end{tabular}
\end{table*}

\begin{figure*}[t!]
\begin{centering}
\includegraphics[width=\textwidth]{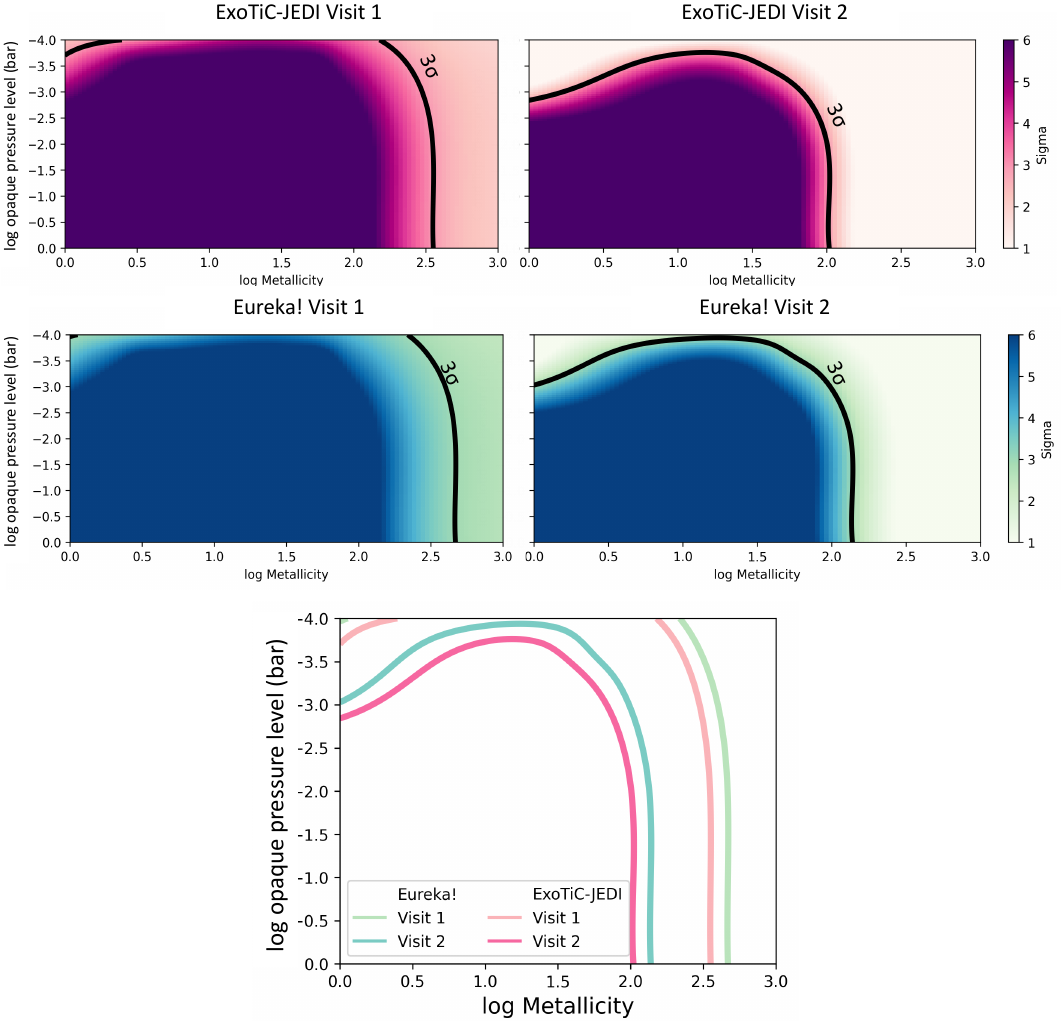}
\caption{Allowed parameter space in terms of opaque pressure level (either a cloud top or surface pressure, bar) versus log metallicity for the \jedi (top, pinks) and \eureka (middle, blues) reductions for Visits 1 (left) and 2 (right). In all cases, we are able to rule out metallicities of at least 100$\times$ solar (i.e., mean molecular weights of $\sim$5.5) with an opaque pressure of 1 millibar by 3$\sigma$. The exact limits of each reduction and visit differ (see Table \ref{tab:fit_table_physical}). Bottom: the 3$\sigma$ limits for each reduction and each visit are shown on the same axis for better comparison.}
\label{figure:heatmaps}   
\end{centering}
\end{figure*}

We additionally interpolate in opaque pressure level vs. metallicity phase space, using a 2D cubic interpolation to compute smooth $\sigma$ contours for our fits, shown in Figure \ref{figure:heatmaps}. We find that the exact cloud-top pressure (or surface pressure) and metallicity we can rule out varies per reduction and per visit. 

For Visit 1, the \jedi data reduction results in pressures greater than 10$^{-3}$ bar ruled out (i.e., discounted by $\geq3\sigma$) up to 350$\times$ solar, while the \eureka reduction can put a stronger upper limit on a 10$^{-3}$ bar pressure level atmosphere, up to 470$\times$ solar metallicity. These correspond to mean molecular weights of 9.9 and 13.7 g/mol, respectively. 
There is a large region of parameter space between 1$\times$ solar and 100$\times$ solar that can be very strongly ($\geq5\sigma$) ruled out at any pressure level tested. As with previous COMPASS targets (e.g., \citealt{Alam2024}, Teske \& Batalha et al., submitted), however, we cannot statistically rule out 1$\times$ solar metallicity atmospheres at lower pressures (10$^{-4}$ bar) for either reduction at 3$\sigma$. The ability to rule out metallicities at $\sim$10$\times$ solar but not 1$\times$ solar at low pressures occurs because of the slight increase in the abundance of methane at mid-level metallicity atmospheres which still have a low mean molecular weight. Thus, these atmospheres have large scale heights combined with a stronger methane feature, which results in overall greater transit depth extents across the G395H bandpass compared to the 1$\times$ solar case at these low pressures. The thin hydrogen-rich atmospheres at low pressure which we are statistically unable to rule out would likely be susceptible to escape and unlikely to persist over long timescales (indeed, such a thin atmosphere would correspond to a mass fraction of $\sim$10$^{-10}$, which would be indistinguishable from a bare rock for photoevaporation models such as \citealt{RogersOwen2021}). At 514\,K and solar metallicity, there are also few cloud or haze species \citep[e.g.,][]{Moran2018, Gao2020} that are likely to exist and build up enough to suppress atmospheric features to this extent. Therefore, of the allowed scenarios for Visit 1, we disfavor the low metallicity, low pressure scenario as being less physically plausible compared to the high metallicity cases.

For Visit 2, we find much more conservative limits on the atmospheric metallicity and opaque pressure level for both reductions. The \jedi reduction rules out pressures greater than 10$^{-3}$ bar up to 100$\times$ solar; the \eureka reduction extends this very slightly to 130$\times$ solar metallicity, with corresponding mean molecular weights of 5.1 and 5.6 g/mol respectively.  Additionally, we lose the ability to rule out even 10$^{-3}$ bar pressures at 1$\times$ solar, compared to the 10$^{-4}$ bar limit found for Visit 1. As with Visit 1, the forward model fits to Visit 2 also includes extended phase space where elevated metallicities at low pressures ($\sim$1 millibar) are ruled out while lower ($\leq$10$\times$ solar) metallicities at the same pressures cannot be. Again, this result occurs at $\sim$10$\times$ solar due to the combination of increased methane abundances while atmospheres are still overall low mean molecular weight ($\sim$2.5\,g/mol compared to 2.3\,g/mol at 1$\times$ solar), producing slightly larger absolute model extents in transit depth. 

The more conservative lower limit on metallicity for Visit 2 in comparison to Visit 1 occurs due to the structure of the transit spectrum across the NRS1 detector. Visit 2 has slightly higher scatter from 3.1 to 3.7\,$\mu$m, as well as a slightly positive slope across the full NRS1 bandpass, which allows methane features in lower metallicity atmospheres to reasonably fit the data, while this region is more uniformly flat in Visit 1, with the overall slope in NRS1 slightly negative. These fitting differences can be seen particularly in the 100$\times$ solar metallicity atmosphere in Figure \ref{figure:forwardmodels} between visits for both reductions.



\section{Discussion and Conclusions} 
\label{section:discussion}
We have presented two JWST NIRSpec/G395H observations of the transmission spectrum of the super-Earth TOI-776b. We produced two independent reductions of the data using the \jedi and \eureka pipelines, resulting in a median transit depth uncertainty of 34\,ppm for both visits and both reductions in 30-pixel wide bins. Though both reductions produce very similar transmission spectra for the same visit, each visit shows a differing overall structure. When fitting for non-physical models, we find that for both reductions, the Visit 1 data prefer a flat line across both detectors while the Visit 2 data prefer a flat line with an offset between NRS1 and NRS2, the size of which is consistent between the two reductions. 

Using \picaso forward models, we can most conservatively rule out atmospheres less than 100$\times$ solar metallicity across all visits and reductions. Beyond this absolute lower metallicity limit, potential atmospheres for TOI-776b include high metallicity atmospheres $\sim$350$\times$ solar in the case of a thin, 10$^{-3}$ bar atmosphere, or as high as 1000$\times$ solar for a 1 bar surface or cloud pressure, though the exact metallicity value below which we can rule out for a specific pressure is dependent on the visit and data reduction in question. 

For Visit 1 in particular, the lower limits we obtain are in fact slightly higher than previous COMPASS inferences, which have tended to lose sensitivity around $\sim$200--300$\times$ solar \citep{Alderson2024,Scarsdale2024,Wallack2024}. We investigated whether these differences could occur, at least in part, due to our slightly differing backend chemistry, as previous COMPASS analyses used the \citet{Lodders2009} solar abundances (M/H=0.0199), compared to our use of the \citet{Asplund2009} solar abundances (M/H=0.0191). However, when we repeated our model fits using the \citet{Lodders2009} abundances, we found that we reproduced nearly exactly the same constraints as with those of \citet{Asplund2009}. This therefore implies that the nature of the noise in our TOI-776b transmission spectra is what may be driving this constraint. This is further borne out in that while the limits we obtain for Visit 1 are at relatively high metallicities, the limits for Visit 2 are more comparable to those for individual visits seen for other planets in the COMPASS program. 

The differences in the metallicity and pressure limits we can obtain per reduction and per visit highlight the care that must be taken when interpreting potential super-Earth and radius valley planetary atmospheres. For Visit 1, the 3$\sigma$ lower limit we obtain at 10$^{-3}$ bar is 350$\times$ solar for \jedi and 470$\times$ solar for \eureka, whereas for Visit 2, \jedi obtains 100$\times$ solar while \eureka obtains 130$\times$ solar (see Figure \ref{figure:heatmaps}). It is interesting that the difference in the lower limit between the \jedi and \eureka reductions is larger for Visit 1 than for Visit 2, despite the fact that the transit depth precisions and difference between the reductions are the same for both visits (see Section \ref{section:results}). In the high metallicity regime that the Visit 1 data supports, forward models for TOI-776b are close to being flat lines, with only small absorption features predominately from methane and carbon dioxide (Figure \ref{figure:forwardmodels}). This is in part due to the fact that at such metallicities, increasing the metallicity further causes the mean molecular weight of the atmosphere to increase, which in turn decreases the scale height of the atmosphere, flattening any atmospheric features. The size of the carbon dioxide and methane features in a 350$\times$ solar model atmosphere compared to the same features in a 450$\times$ solar atmosphere for TOI-776b differ by only 5--7\,ppm. The \jedi and \eureka reductions differ by a median of 8\,ppm in the region centered approximately on the methane band (3.2--3.6\,$\micron$) and by 12\,ppm centered approximately on the carbon dioxide band (4.15--4.55\,$\micron$) in Visit 1. These differences are considerably smaller than the corresponding transit depth precisions ($\sim$26\,ppm and $\sim$43\,ppm respectively), which would suggest that the reductions are consistent given the uncertainty of the data, however, they are comparable in size to the difference in absorption feature sizes between the 350$\times$ and 450$\times$ solar models. This indicates how the two reductions can obtain differing lower limits on the metallicity, and demonstrates that relatively large differences in quoted metallicity lower limits can arise due to even $\sim$ppm differences in the per-point transit depth between reductions.


Our results provide a note of caution in over-interpreting any single stated metallicity or pressure level constraint for super-Earth atmospheres, as the swath of allowed atmospheres clearly varies by visit and reduction. Particularly when comparing models that differ only by 10s of ppm against transit depth precisions of similar scale, minor differences can result in wide variations in interpretation. This will become particularly important as we continue to build up a library of terrestrial/super-Earth atmospheric observations, compare results across large programs (e.g., GO-1981; GO-4098; this program, GO-2512), and attempt to discern a coherent picture for the broader super-Earth population. Looking ahead to the planned Director's Discretionary Time (DDT) program \citep{Redfield2024} to constrain terrestrial atmosphere emission, similar care should be taken in comparing across modeling tools and interpretations.

In returning to our interpretation of TOI-776b, our non-physical models resulted in flat (or flat accounting for an offset as in Visit 2) models as the best fits, consistent with our physical modeling, which results in the lowest $\chi^2/N$ for very flat transmission spectra due to highly enhanced metallicities at 1 bar ($\sim$1000$\times$ solar) or a very thin atmosphere (i.e., pressures of 10$^{-4}$ bar). While no atmosphere at all is also possible given our modeling, the density of TOI-776b, along with its position within the radius valley, is difficult to reconcile with a bare rock devoid of some lower-density material (see e.g., Figure 1 in Teske \& Batalha et al., submitted). Taken together, our non-physical and physically consistent models suggest that the most likely explanation for our observations of TOI-776b is that the planet either has a very thin, very cloudy, or highly metal-rich atmosphere. \linebreak







\noindent The data products for this manuscript can be found at the following Zenodo repository: 10.5281/zenodo.14720057.
We thank the anonymous referee for their swift review.
This work is based on observations made with the NASA/ESA/CSA James Webb Space Telescope. The data were obtained from the Mikulski Archive for Space Telescopes at the Space Telescope Science Institute, which is operated by the Association of Universities for Research in Astronomy, Inc., under NASA contract NAS 5-03127 for JWST. These observations are associated with program \#2512. Support for program \#2512 was provided by NASA through a grant from the Space Telescope Science Institute, which is operated by the Association of Universities for Research in Astronomy, Inc., under NASA contract NAS 5-03127.
L.A. acknowledges funding from UKRI STFC Consolidated Grant ST/V000454/1 and is supported by the Klarman Fellowship.
S.E.M. is supported by NASA through the NASA Hubble Fellowship grant HST-HF2-51563 awarded by the Space Telescope Science Institute, which is operated by the Association of Universities for Research in Astronomy, Inc., for NASA, under contract NAS5-26555.
This work is funded in part by the Alfred P. Sloan Foundation under grant G202114194.
Support for this work was provided by NASA through grant 80NSSC19K0290 to J.T. and N.W. 
H.R.W. was funded by UK Research and Innovation (UKRI) under the UK government’s Horizon Europe funding guarantee [grant number EP/Y006313/1].
This material is based upon work supported by NASA’S Interdisciplinary Consortia for Astrobiology Research (NNH19ZDA001N-ICAR) under award number 19-ICAR19\_2-0041.
This work benefited from the 2023 Exoplanet Summer Program in the Other Worlds Laboratory (OWL) at the University of California, Santa Cruz, a program funded by the Heising-Simons Foundation. 
This research made use of the NASA Exoplanet Archive, which is operated by the California Institute of Technology, under contract with the National Aeronautics and Space Administration under the Exoplanet Exploration Program.

Co-Author contributions are as follows: 
LA led the data analysis and write-up of this study. SEM led the modeling efforts with major contributions from NEB and NW. NLW provided reductions and analyses of the data. AD provided expertise regarding photoevaporation modeling.
HRW advised throughout the analysis. All authors read and provided comments and conversations that greatly improved the quality of the manuscript.

\software{\texttt{astropy} \citep{Astropy2013, astropy, AstropyCollaboration2022}, \texttt{batman} \citep{Kreidberg2015}, \texttt{Cantera} \citep{Goodwin2022}, \texttt{emcee} \citep{Foreman-Mackey2013}, \texttt{Eureka!} \citep{Bell2022},  \texttt{ExoTiC-JEDI} \citep{Alderson2022}, \texttt{ExoTiC-LD} \citep{Grant2022}, \texttt{Matplotlib} \citep{matplotlib},  \texttt{NumPy} \citep{numpy}, \texttt{pandas} \citep{pandas}, \texttt{PandExo} \citep{Batalha2017}, \texttt{PICASO} \citep{Batalha2018, Mukherjee2023}, \texttt{photochem} \citep{Wogan2023}, \texttt{scipy} \citep{scipy}, STScI JWST Calibration Pipeline \citep{Bushouse2022}, \texttt{ultranest} \citep{Ultranest}, \texttt{xarray} \citep{xarray}} 

\facilities{JWST (NIRSpec)} 

The JWST data presented in this paper were obtained from the Mikulski Archive for Space Telescopes (MAST) at the Space Telescope Science Institute. The specific observations analyzed can be accessed via DOI: 10.17909/tdmn-7518.

\bibliography{refs}{}
\bibliographystyle{aasjournal}

\newpage
\appendix

\vspace{-3em}

\begin{figure*}[!hb]
\begin{centering}
\includegraphics[width=0.94\textwidth]{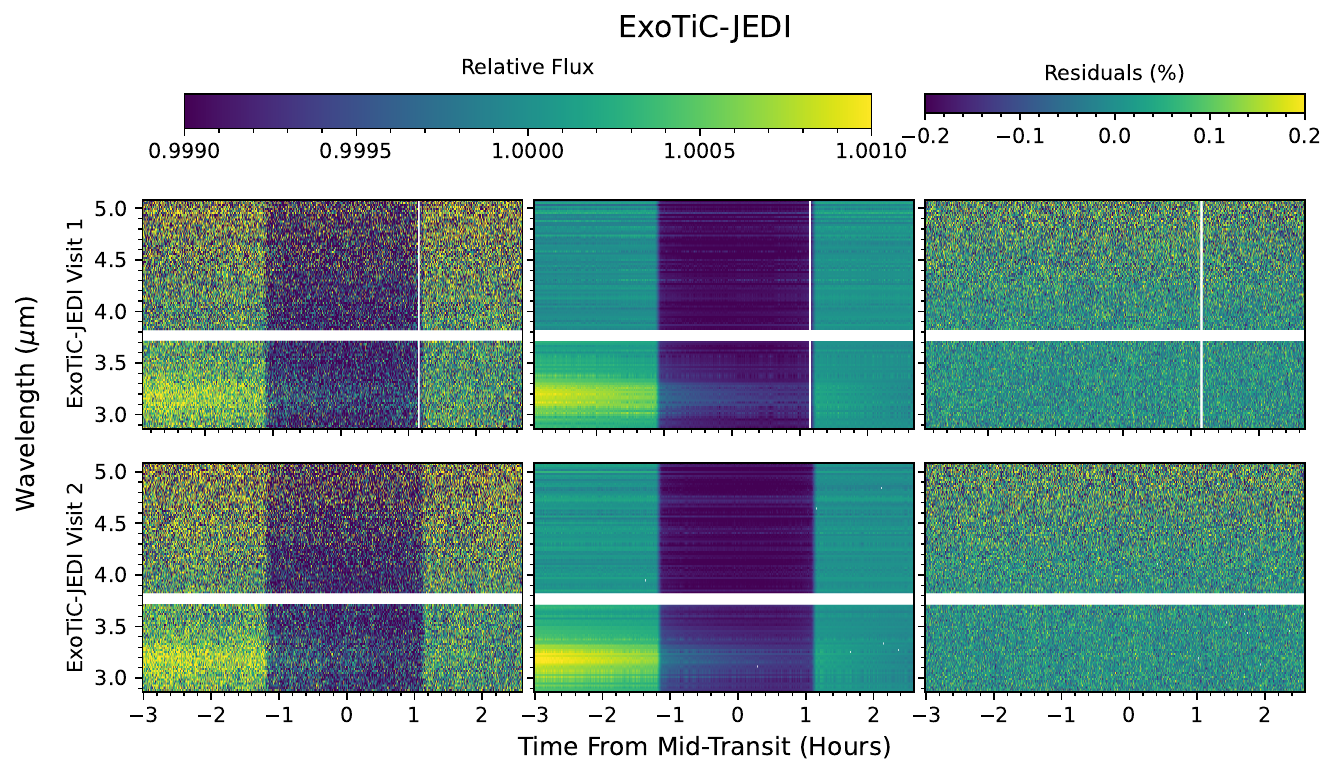}
\includegraphics[width=0.94\textwidth]{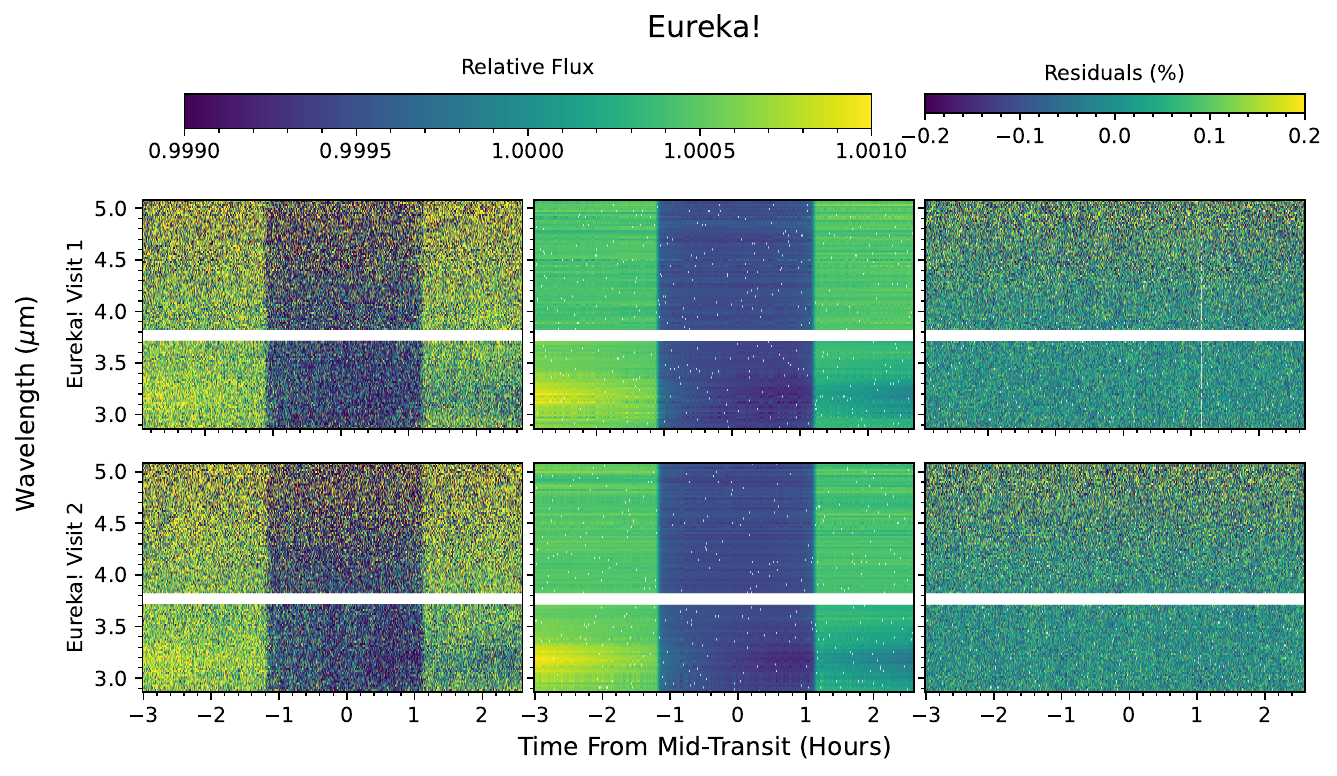}
\caption{2D light curves for \jedi and \eureka for Visit 1 and Visit 2. Columns show the data (left), models (middle), and residuals (right) at $R\sim200$, the resolution plotted in Figure \ref{figure:tspec} and used in the modeling and interpretation. The white vertical stripe in the \jedi Visit 1 data corresponds with the observed HGA move, where 15 integrations were removed from the time series. The \eureka reduction elected not to manually remove any integrations surrounding the HGA move, instead relying on light curve level outlier rejection.} 
\label{figure:2d_plot}   
\end{centering}
\end{figure*}

\end{document}